# A subpicotesla diamond magnetometer

Thomas Wolf[1)*)], Philipp Neumann[1)*)], Kazuo Nakamura[2)], Hitoshi Sumiya[3)], Junichi Isoya[4)]

and Jörg Wrachtrup[1)]

1) 3rd Institute of Physics and Research Center SCoPE, University of Stuttgart, 70550 Stuttgart, Germany
2) Energy System Research Institute, Tokyo Gas Co., Ltd., Yokohama 230-0045 Japan
3) Sumitomo Electric Industries, Ltd., Itami, Hyogo, 664-0016 Japan
4) Research Center for Knowledge Communities, University of Tsukuba, 1-2 Kasuga, Tsukuba, Ibaraki 305-8550 Japan

**NV defect centres in diamond are promising solid-state magnetometers. Single centres allow for high spatial resolution field imaging but are limited in their magnetic field sensitivity to around $10\ \text{nT}/\sqrt{\text{Hz}}$ at room temperature. Using defect centre ensembles sensitivity can be scaled as $\sqrt{N}$ when $N$ is the number of defects. In the present work, we use an ensemble of $10^{11}$ defect centres for sensing. By carefully eliminating all noise sources like laser intensity fluctuations, microwave amplitude and phase noise we achieve a photon shot noise limited field sensitivity of $0.9\ \text{pT}/\sqrt{\text{Hz}}$ at room-temperature with an effective sensor volume of $8.5\text{e}-4\ \text{mm}^3$. The smallest field we measured with our device is $100\ \text{fT}$. While this denotes the best diamond magnetometer sensitivity so far, further improvements using decoupling sequences and material optimization could lead to $\text{fT}/\sqrt{\text{Hz}}$ sensitivity.**

*) Corresponding authors

TW (email: t.wolf@physik.uni-stuttgart.de), PN (email: p.neumann@physik.uni-stuttgart.de)

Magnetic sensors find application in various areas of science and technology[1,2]. Persistent efforts have led to the development of new, highly sensitive magnetic sensors[3,4] as well as the improvement of existing technologies[5]. Improving sensitivity has been a strong motivation for development of subfemtotesla magnetometers[4]. However, due to the $1/r^3$ decay of magnetic dipolar fields, sensor size is a critical further parameter. Consequently, a number of approaches are striving for high sensitivity in combination with reduced sensor sizes[6–9]. Figure 1d compares magnetic field sensitivities and characteristic sizes for various implementations restricted to room-temperature sample and far field techniques. In essence, the graph shows that small sensors with subpicotesla sensitivity have not been realized so far.

The favourable material properties of diamond as well as the optical and spin properties of nitrogen vacancy (NV) defect centres allow for optical polarization, manipulation and readout of its spin state[10]. This opens new ways for the implementation of robust solid state sensors for a variety of quantities[11,12]. In particular as magnetic field sensors, NV-based approaches offer opportunity for detection of magnetic field signals both with high spatial accuracy (nanometer) as well as high field sensitivity[9,13]. In this work we focus on enhancing sensitivity of magnetic field measurements with ensembles of NV centers[14]. While this approach sacrifices the potential atomic scale resolution of single spin magnetometers it has the potential of gaining higher field sensitivity with still smaller sensor dimensions than e.g. atomic vapour-based designs. Magnetic field detection is based on ground state Zeeman shifts of spin sublevels of NV centres $\Delta E = \gamma \hbar B$, where $\gamma$ is the gyromagnetic ratio of the electron spin and $B$ is the field to be measured. $\Delta E$ is best determined by exploiting coherent control of the electronic spin state of the NV centres in its ground state (Fig. 1a-b). In essence the spin acquires a phase $\varphi = \gamma \cdot B \cdot T_\varphi$ during sensing time $T_\varphi$ ($B$ is the averaged field) in Ramsey- or spin echo-type measurements[15]. Optical excitation with a laser pulse concludes a single field evaluation step by invoking spin state dependent fluorescence and reinitializing the spin state via the spin selective singlet decay of the NV centers.[10] In essence, the fluorescence response of the system $S$ is modulated with $\sin \varphi(B)$.

In general, the sensitivity of a magnetic field measurement is given by $B_{min}(t) = \sigma(t)/(\mathrm{d}S/\mathrm{d}B)$, where the standard deviation of the sensor's signal $\sigma(t)$ is compared to the response of the system $\mathrm{d}S$ in a changing magnetic field $\mathrm{d}B$. For the particular case of NV centres using a pulsed detection scheme with discrete readout steps the sensitivity is written as

$$B_{min}(t) = \frac{\sigma_1}{\gamma \cdot A \cdot T_\varphi \cdot \sqrt{n}} \qquad (1)$$

Here $n = t/T_{seq}$ is the number of field evaluations for a total measurement time $t$ with $T_{seq}$, $\sigma_1$ and $T_\varphi$ the duration, the standard deviation and the phase accumulation time of a single field evaluation, respectively. Parameter $A$ is the system-specific amplitude of the signal modulation[15].

Before dwelling on the accuracy of ensemble magnetometry it is instructive to analyse single spin measurements. The standard deviation of single spin sensor readouts $\sigma_1$ is dominated by shot noise of the fluorescence signal (essentially projection noise in photon number). Its ultimate limit however is spin projection noise due to the statistical nature of the quantum mechanical read out of the spin state. It is only reached by reducing the relative fluorescence shot noise below the spin projection noise limit. Steps towards this goal are for instance improved fluorescence detection efficiency by wave guiding effects as shown in Ref.[16], repetitive readout[17,18] or generally different detection schemes[19,20]. Since both, fluorescence signal and spin projection, are sources of uncorrelated noise, sensitivity scales with $\sqrt{n}$ the number of single sensor readouts over a wide range of measurement times.

We now turn to ensemble magnetometry. To further improve sensitivity, $\sigma_1$ is decreased with increasing fluorescence signal intensity when measuring on ensembles of NV centres. For independent emitters $\sigma_1$ should scale as $1/\sqrt{N}$ where $N$ is the number of defects contributing to field measurement. Eventually, we calculate the spin projection limited magnetic field sensitivity as

$$B_{QPN}(t) = \frac{1}{\gamma\sqrt{N \cdot t/T_{seq}} \cdot T_\varphi \cdot e^{-\delta(T_\varphi)}} \qquad (2)$$

with $e^{-\delta(T_\varphi)}$ describing the decay of spin coherence and $N$ the number of NV centers contributing to the signal. The equation is equivalent to the general derivation of Chin et al.[21] for the case $T_{seq} \to T_\varphi$. In this case and for an exponential decay of spin coherence with a time constant $T_2$, the minimum of equation (2) is achieved for $T_\varphi = T_2/2$, which simplifies to $B_{QPN}(t) = \frac{\sqrt{2e}}{\gamma\sqrt{NtT_2}} = 1.3 \cdot 10^{-11}$ T/$\sqrt{NT_2}$.

Our sensor consists of a 0.9 ppm NV⁻ HPHT-diamond, starting from 3ppm nitrogen before conversion to NV by electron irradiation and has an optical thickness of 500 µm with <111> front planes. Fluorescence from the NV centres after pulsed excitation using a green laser is measured on one channel of a difference detector. The second channel of the detector is illuminated by part of the green excitation beam split from the exciting laser (Fig. 1a). Measurements shown were conducted at room temperature with a constant offset magnetic field of 46 Gauss along one NV orientation used for the measurements. The sensitivity of single centre magnetometry is limited by the number of detectable photons from a single defect being given by the photo-physics of the defect. In ensemble magnetometry, an additional challenge is to detect as much fluorescence photons as possible from a finite sample volume. As a result, different collection as well as absorption schemes have been proposed[16,17,19,20,22]. In our approach, we maximize fluorescence signal intensity from the NV centres by using a parabolic shaped glass lens contacting one side of the diamond. As the étendue of a light source is conserved (for constant intensity), this structure essentially trades the initially large solid angle of fluorescence radiation against size of the emitting surface area of the structure. Simulations of the structure using commercial software show a collection efficiency higher than 60% without attempts for geometrical optimization of the diamond (Fig. 1c). Hence, the structure offers a collection efficiency comparable to the previously reported side detection approach[16] with the additional advantage of a directed fluorescence output behind the structure. As a result, only one detector is needed for detection of the signal.

Our magnetic field measurement scheme comprises three steps. First, the NV sensor spins are polarised with a laser pulse. After initialization, we apply a microwave preparation sequence for B field measurement. The fluorescence signal is triggered and read out subsequently by launching another laser pulse. Microwave pulses are implemented using a coil antenna. Typical Rabi frequencies are on the order of 5 MHz.

For excitation, we focus 400 mW of laser power with a diameter of 47µm onto the sample. The maximum intensity used is 25 kW/cm², which is below saturation (~100 kW/cm²). The sample volume and hence the number $N$ of defects contributing to the fluorescence signal is determined by the optical excitation and detection volume. Based on a measurement of the excitation area using a CCD camera and given the collection property of our parabolic lens, which is in first order spatially non-selective towards the fluorescence created, we calculate a detection volume of $V = 8.5e-4$ mm³. With a density of 0.9 ppm we estimate that 1.4e11 NV centres contribute to the sensor signal.

From equation (1) we estimate an expected maximum sensitivity with $T_\varphi = 50 \mu s$ of $100\, fT/Hz^{1/2}$ if we scale values of single NV sensitivities with the mentioned number of NV centres. From equation (2) we calculate the spin projection noise limit to be $6\, fT/Hz^{1/2}$ ($T_\varphi$ is again set to 50μs). This estimate relies on the assumption that the results of single readout steps show a normal distribution around a well-defined value (central limit theorem). This condition is usually met for measurements on single NV centres with comparably small numbers of total signal photons dominated by optical shot noise or spin projection noise – a frequency independent, uncorrelated white noise background. Ensemble magnetometry, however, dealing with much higher fluorescence intensities, is plagued by other, correlated and time-dependent noise sources. Since preparation and readout of the measurement relies on discrete preparation steps using laser and microwave, it is essential to analyse to what degree each of these sources influences the sensitivity and how to mitigate their impact.

We start by performing an AC-magnetometry experiment as demonstrated previously[14,23]. To this end, a spin-echo measurement with pulses $(\pi/2)_x - (\pi)_x - (\pi/2)_y$ is phase-locked to a sinusoidal ac magnetic field (see figure 2a and methods), which we intend to sense. We use a phase accumulation time of $T_\varphi = 50$ μs and the overall single sequence length is $T_{seq} = 160$ μs. The phase shift of the last π/2 pulse in the echo sequence assures maximum sensitivity already for smallest amplitude of the test field. By increasing the ac amplitude, we increase the accumulated phase linearly with a concomitant sinusoidal fluorescence response. For the following investigations on the reproducibility of single sensor readouts and its scaling behaviour with averaging time, we chose the point of maximum field sensitivity.

Figure 2b shows the scaling of the Allan deviation of the readout signal of two different measurements. The upper curve corresponds to the spin-echo measurement described above. The second stems from an identical measurement but without applying microwave pulses in between laser readout pulses. While the Allan deviation may not be considered a valid estimator for the scaling of magnetic field sensitivity, it provides information on the correlation of consecutive measurements (here $t_{corr} < 100$ ms) without being affected by overall (long-term) drift as in case of the standard deviation. As can be seen in the graph, once microwave pulses are applied sensitivity does scale worse than $\sqrt{n}$. No further improvement by averaging is achieved. Effects related to the implementation of the microwave sequence prevent favourable scaling of the readout signal. Note that the Allan deviation in conjunction with the applied spin echo sequence on short timescales suppresses effects of long-term drifts e.g. due to changes in temperature[11] or magnetic background field. Thus, only external magnetic noise and variations of the implemented microwave sequence in the frequency range shown remain obvious culprits.

The impact of microwave amplitude and frequency noise on measurement error is in general dependent on the particular choice of the microwave pulse sequence and noise frequency. For noise correlation times longer than the length of a single readout sequence ($T_{seq}$) an error in microwave amplitude or frequency can be taken to be constant throughout a single sequence. In this regime the signal error after a microwave pulse sequence due to a set of amplitude and frequency error parameters can be extracted from simulations of coherent spin rotations under the NV spin Hamiltonian[24,25]. Figure 2c shows results for the scaling of signal error for two limiting cases: 1) Scaling with the relative microwave power error *Δg* (*Δf*=0) and 2) Scaling with the absolute microwave frequency error *Δf* (*Δg*=0). Both quantities show a linear scaling behaviour in the relevant parameter range.

We measured amplitude and phase noise of the microwave system using established cross-correlation techniques. Both are given for frequencies *f* below the inverse sequence length ($1/T_{seq}$) in figure 2d (cumulative root sum of squares from $1/T_{seq}$ to $f$). The optical noise is given in the same way and the Allan deviation of single sequence readouts is indicated as horizontal black line. Three results can be inferred from the graph. First, the influence of microwave frequency noise on scaling of the signal is negligible. Secondly, the optical low frequency noise ($f < 1/T_{seq}$) is well below the mutual deviation range of single sequence readouts. This enables the preferential scaling with $\sqrt{n}$ for the case where no microwave pulses were applied. Finally, the effect of microwave amplitude noise is small on the timescale of single readout steps but increases towards longer timescales and exceeds the level of deviation of single readout steps. In the latter case, scaling of the measurement is worse than $\sqrt{n}$ as the central limit theorem does not apply. The centre of the distribution of measurement values rather drifts around a new, larger range now being dominated by microwave amplitude noise. In essence, varying microwave amplitudes lead to improper conversion of field to signal amplitude that prevents a decrease of the measurement error with increasing averaging time.

One way to reduce the impact of inaccurate microwave pulses on scaling of the readout signal is to reference the signal on a timescale shorter than the characteristic correlation time of the noise. Owing to the photo-physical dynamics of the NV centre, the spin signal is typically read out in the first part of a laser pulse. It is common practice to reference this signal ('1' in figure 3a) to the steady state fluorescence level after re-initialization of the NV centres at the end of the laser pulse ('2' in figure 3a) yielding the measurement signal $S_B$. Implicitly, this procedure mitigates optical noise with a correlation time longer than the laser pulse length (100μs). Effectively, it implements a filter $X_B$ for optical noise frequencies lower than the inverse laser pulse length, i.e. 10 kHz. The filter $X_B$ related to signal $S_B$ can be calculated by the Fourier transform of the respective signal integration window $C_B$ by

$$X_B(\omega) = \left|\int_{-\infty}^{\infty} dt\, e^{i\omega t} C_B(t)\right| \quad (3a)$$

$$C_B(t) = \begin{cases} 1, & t \in [0, \Delta t] \\ -1, & t \in [t_L - \Delta t, t_L] \\ 0, & \text{otherwise} \end{cases} \quad (3b)$$

with the length of the laser pulse $t_L$ and the integration time $\Delta t$ (see also figure 3a). We stress that this filter is different from 1) filter functions implemented with microwave pulses with the intention to shape the response of NV centres towards a certain band of magnetic signal frequencies[26,27] and 2) filter functions related to data (post-) processing (e.g. to improve the signal-to-noise-ratio). The filter referred to here is an intrinsic part of the measurement system based on the fact that we measure discrete, non-continuous time frames.

While the described filter only affects the laser induced correlated fluctuations of the signal, a reference for the state preparation with microwave pulses can be established likewise by introducing a second preparation and readout sequence as shown in figure 3a for signal $S_D$. To understand the impact of this procedure we calculate the filter transmission for the four different ways of measuring the NV signal ($S_A$ to $S_D$) shown in figure 3a. Signal $S_A$ does not contain any referencing. $S_B$, as explained above, implements one filtering step for the optical signal on the timescale of the laser pulse length. $S_C$ results in one referencing step for both optical and microwave-related noise on the timescale of the sequence length. Finally $S_D$ gives two referencing steps versus the optical part and one for the microwave contribution.

Calculation of filters $X_{A/C/D}$ is analogous to $X_B$ and the calculated filter responses are given in figure 3b for the same set of parameters used for the measurements from above $t_L = 100$ μs, $\Delta t = 10$ μs and $T_{seq} = 160$ μs. The calculation of the filter transmission is herein restricted to noise frequencies in the low frequency regime up to the range of $1/T_{seq}$. Note that the response of the measured signal corresponding to the implementation of the signal measurement procedure $S_D$ results in filter $X_D$ for optical noise because of two referencing steps. It results in filter $X_C$ for the microwave noise contributions since state preparation by microwave only affects the signal in the first part of the laser pulses.

With the calculated filter responses, we again give the cumulative noise (root sum of squares) weighted with the respective filters for microwave components and the optical part corresponding to the measurement procedure $S_D$ from two consecutive readouts (Fig. 3c). The calculation shows that referencing the signal in this way very efficiently suppresses low frequency noise components. The contribution of microwave amplitude noise is kept below the limit set by the deviation of single readout steps.

Next we repeat the ac-magnetometry sequence as described above and in figure 2a, however, measuring signals $S_B$ and $S_D$. After varying the strength of the magnetic test field to retrieve the amplitude of the sensor response (figure 3d) we again switch off the actual ac field to 1) reach the working point of highest sensitivity and 2) exclude additional noise sources from the ac signal itself. The scaling of the sensitivity, given here as the standard deviation for the two signals $S_B$ and $S_D$ is shown in figure 3e. In case of signal $S_B$ (blue) increased averaging time does not improve the measurement result. Signal $S_D$ (green) shows a scaling with $1/\sqrt{t}$ where t is the total signal averaging time and reaches a sensitivity of $0.9\ \text{pT}/\text{Hz}^{1/2}$. For longer measurement time an absolute sensitivity of around 100 fT is achieved.

The sensitivity shown marks an improvement by three orders of magnitude in magnetic field sensitivity when compared to previously published NV diamond related results[16]. Secondly, we show for the first time a $\sqrt{t}$ scaling behaviour in NV ensemble magnetometry, being a requirement for strategies to improve sensitivity. Moreover, the scaling behaviour shows that the measurement is so far limited neither by temperature variation nor by external in-band magnetic noise. We find that the standard deviation of the signal $S_D$ is by a factor of 5.3 above the fluorescence shot noise level, that we would expect for a simple readout of the repolarization signal only ($S_A$). This agrees well with the expected increase in uncorrelated noise. While correlated noise is largely suppressed with the procedures described, uncorrelated noise increases by a factor $\sqrt{2}$ with every referencing step implemented if we assume identical noise density for the uncorrelated noise of the two signals referenced. Since we introduced 3 referencing steps (exciting laser against fluorescence and two in measurement procedure $S_D$) and an additional factor for doubling the measurement time, we thereby effectively increase the contribution of uncorrelated noise by $\sqrt{2}^4 = 4$. Concluding, we implemented a self-referencing measurement of a single sensor at different times. Instead mutual referencing of two sensors at the same time could yield an improvement of $\sqrt{2}^3$.

When compared to the sensitivity extrapolation from measurements on single NV centres from equation (1), we find a deviation by one order of magnitude. We can resolve this discrepancy by accounting for the reduction in contrast when measuring on one of four NV-axes and the increase in uncorrelated noise by 5.3 mentioned before. We predict that the sensitivity of measurements on NV ensembles can even exceed the projection derived from single NV measurements due to improved fluorescence collection efficiency. Finally, we want to emphasize that the implementation of the measurement procedure $S_D$ except for a decrease in measurement rate does not impose any additional restrictions. In particular, we remark that, as we require control of the ac-magnetic field (e.g. switch off in every second measurement, in-phase with spin echo sequence), also in real measurements the source of ac magnetic field needs to be controlled (e.g. invoked flips of electron or nuclear spins to be measured[28,29]).

The present work highlights the role of technical noise and its mitigation to sensitivity scaling in NV ensemble magnetometry. By reducing the influence of non-white noise contributions over an extended frequency range we achieve a $\sqrt{t}$ scaling in sensitivity close to the photon shot noise limit, finally reaching sub pT/$\sqrt{\text{Hz}}$ sensitivity for a sensor volume of 8.5e– 4 mm$^3$. Recurring to figure 1d, this places our current sensor in terms of sensitivity per volume among state-of-the-art sensor implementations. Different strategies are conceivable to further improve the magnetic sensitivity. In Ref.[30] higher order dynamical decoupling sequences were applied to NV ensembles yielding a phase memory time of 2 ms, which is the limit set by longitudinal spin relaxation of NV centres at room temperature. With the experimental settings described here, with a similar amount of NV centres and identical efficiency of the filters applied this would yield a sensitivity of 40 fT/$\sqrt{\text{Hz}}$ . The latter is still almost two orders of magnitude above the limit set by the spin projection noise (0.9 fT/$\sqrt{\text{Hz}}$). This value itself allows for detection of proton spins in a microscopically resolvable volume in less than one second. Nuclear spin assisted repetitive readout[17], infrared absorption based readout[19] or enhancement by optical cavities[20] are strategies to reach the projection noise limit.

**Methods:**

Allan deviation:

The (non-overlapping) Allan deviation $\sigma_A(\tau)$ of a set of data samples $S = [S_1, S_2, \ldots, S_n]$ with sample spacing $t'$ is defined for a given time interval $\tau$ by:

$$\sigma_A^2(\tau) = \frac{1}{2} \langle (x_{i+1} - x_i)^2 \rangle_\tau \qquad (4)$$

Here, $x_i$ denotes the mean ($\langle \cdot \rangle$) over the subset of $m = \tau/t'$ successive elements of $S$ within the i$^{\text{th}}$ $\tau$-interval:

$$x_i = \langle [S_{(i-1)m+1}, S_{(i-1)m+2}, \ldots, S_{i \cdot m}] \rangle \qquad (5)$$

AC magnetometry sequence:

For AC magnetometry we apply the most basic sequence, namely a Hahn echo measurement (($\pi/2$) – ($\pi$)– ($\pi/2$)). Hence, a microwave $\pi/2$ pulse creates a spin superposition state followed by two equal free evolution times $T_\varphi/2$ separated by a $\pi$ pulse. The AC signal has to have the frequency $1/T_\varphi$ and has to be in phase with the $\pi$ pulse (e.g. the zero crossing of a sine wave has to coincide with the $\pi$ pulse) in order to yield highest field sensitivity. The accumulated phase of the sensing spins is proportional to the field strength ($\varphi = \gamma \cdot B \cdot T_\varphi$). Finally, a second $\pi/2$ pulse is applied to convert phase into a detectable spin population difference (e.g. the population of spin projection $m_S = 0$, $p_{m_S=0} = \frac{1}{2}(1 + \cos \varphi)$). Highest sensitivity is achieved around the point of equal spin state population (i.e. $p_{m_S=0} = 0.5$). Adjusting the phase $\phi$ of the final microwave pulse assures the optimal working point for arbitrary field strengths (i.e. $p_{m_S=0} = \frac{1}{2}(1 + \cos(\varphi + \phi))$). Consequently, for highest sensitivity to magnetic fields around zero amplitude the final pulse has to be phase shifted by $\phi = 90°$. Thus, our Hahn echo measurement sequence changes to $(\pi/2)_x$ – $(\pi)_x$ – $(\pi/2)_y$. In addition, high dynamic range magnetometry can be applied to remove field ambiguities and at the same time retain highest sensitivity.[31,32]

Error scaling with microwave amplitude and frequency:

In order to estimate the impact of microwave pulse errors on the measurement we calculate the population difference between the target state (ideal pulses) and the outcome of a pulse sequence with constant error in microwave frequency and microwave power throughout a sequence (Hahn echo). Successive coherent spin rotations are calculated using the NV spin Hamiltonian:

$$\mathcal{H} = DS_z^2 + B_z(\gamma S_z + \gamma_n I_z) + AS_z I_z \qquad (6)$$

$D = 2.87$ GHz is the zero-field splitting, $\gamma/2\pi = 28.7$ GHz/T is the gyromagnetic ratio of the NV-electron spin, $\gamma_n/2\pi = 3.08$ MHz/T the nuclear gyromagnetic ratio of $^{14}$N and $A = 2.16$ MHz is the hyperfine coupling between NV-electron and $^{14}$N-nuclear spin. $S_z$ and $I_z$ are the electron and nuclear spin projection operators respectively.

Filter Functions:

Explicit evaluation of equations 3a-b yields the filter function $X_B$:

$$X_B = \sqrt{|2/\omega^2 \cdot [(2 - 2 \cdot \cos \omega \Delta t + \cos \omega(t_L - 2\Delta t) + \cos \omega t_L - 2 \cdot \cos \omega(t_L - \Delta t)]|}$$

Filter functions $X_{A/C/D}$ are calculated in an analogous way using the corresponding signal integration windows $C_{A/C/D}$.

**Acknowledgements:**

We thank Bernhard Grotz, Marco Di Sarno and Steffen Steinert for fruitful discussions. We acknowledge financial support by the German Science Foundation via SFB/TRR 21, FOR 1493 and via the Japanese-German (JST, DFG) joint research group FOR 1482. Furthermore, we received funding from the European Union via SIQS and ERC SQUTEC and from the Max-Planck-Society.

**Author Contributions:**

JI, HS and KN synthesized and characterized the NV diamond sample, TW performed the experiments. All authors designed the experiments, discussed the results and wrote the paper.

**Competing Financial Interests:**

The authors declare no competing financial interests.

Figures:

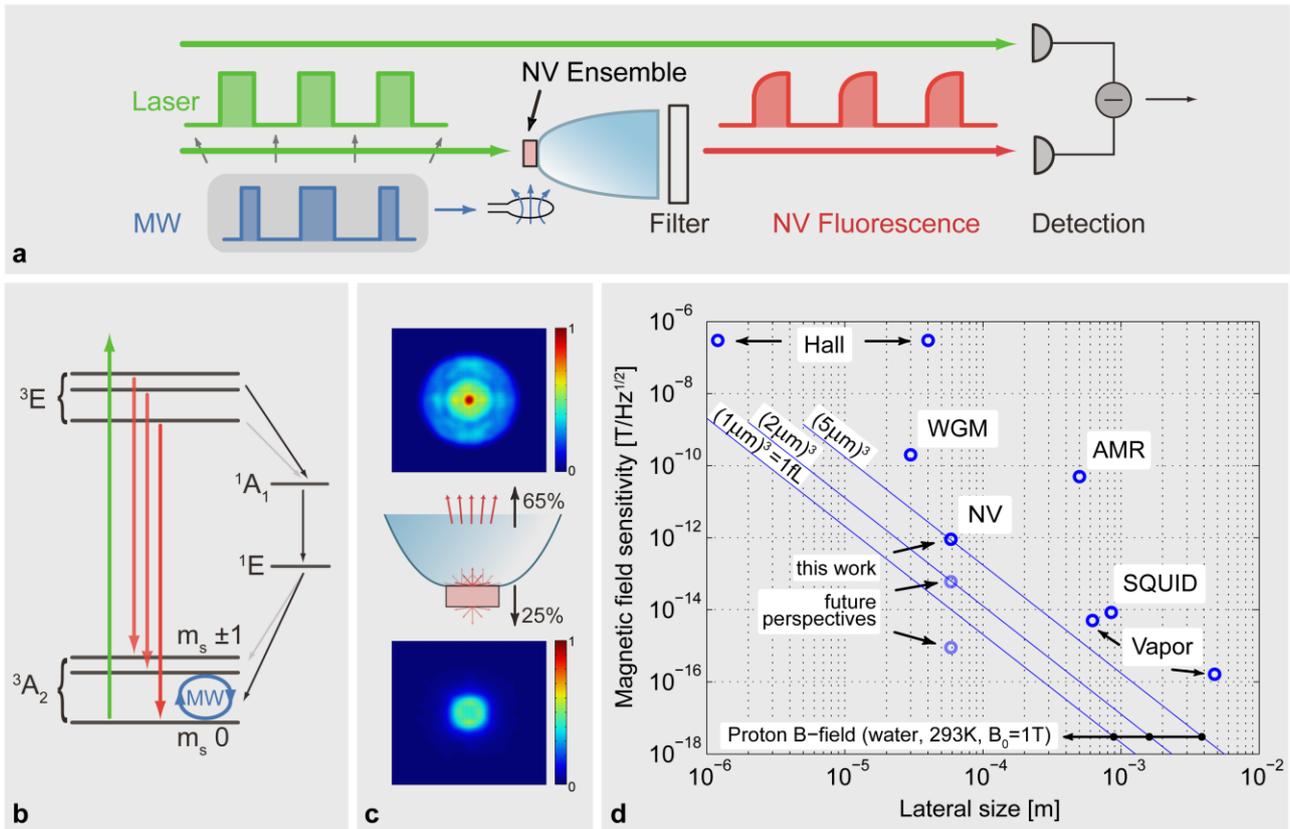

**Fig. 1: NV diamond ensemble magnetometer. a**, Pulsed NV experiment: NV ensemble is excited by 532nm laser pulses. Long-pass filtered fluorescence is collected with part of the exciting light on a difference photo detector. Microwave (MW) is used for NV spin manipulation. **b**, NV energy level scheme: Manipulation of electron spin in triplet ground state. Spin state dependent fluorescence allows read out of the spin state. **c**, Fluorescence collection with parabolic collector (simulation results). **d**, Magnetic field sensitivity versus sensor-to-sample distance – comparison chart including various available sensor techniques with sample at room-temperature and remote detection: vapour cells (optical readout)[4,33], inductively coupled SQUID[3], anisotropic magnetoresistance sensor (AMR)[5], whispering gallery mode resonator based sensor (WGM)[34], Hall-sensors[5] and NV ensemble sensor (this work (upper point) and future perspectives (middle point: $T_\varphi$=2 ms, lower point: spin projection noise limit with $T_\varphi$=2 ms). X-coordinate in all cases given as the radius of an assumed spherical detection volume calculated from stated volumes/sensor size in references. For comparison straight lines indicate for various volumes the magnetic field from nuclear spins of protons in water at room-temperature and external magnetic field $B_0$=1T. Lowest line corresponds to the resolvable volume in optical confocal microscopy (1 femtoliter).

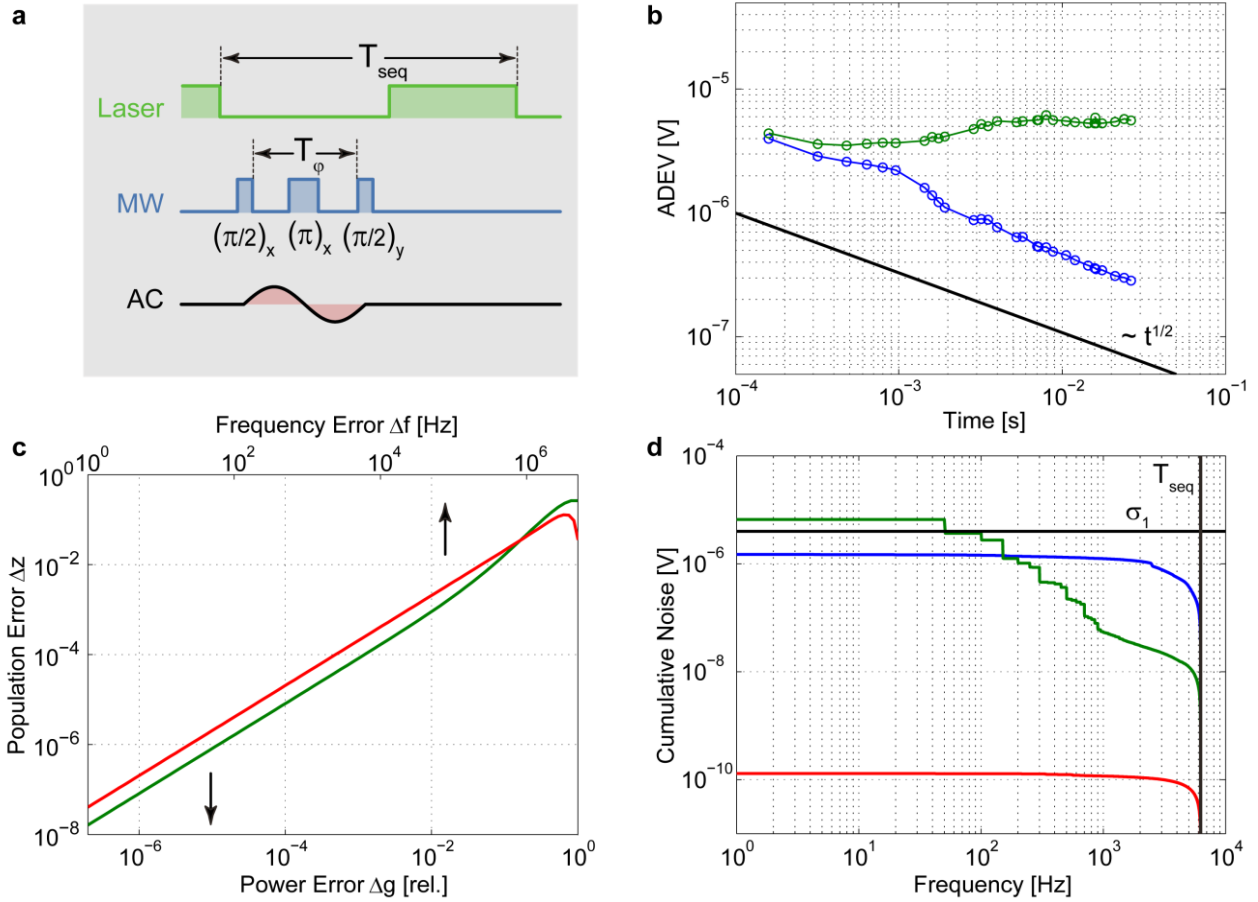

**Fig. 2: Influence of non-white noise on magnetometer sensitivity. a**, AC magnetic field measurement scheme with pulsed sensor readout. **b**, Scaling of Allan deviation from Hahn Echo sequence with (green) and without (blue) microwave pulses. Slope of black line indicates the desired scaling behaviour to approach a central limit. In case microwave pulses are applied scaling is worse than $\sqrt{t}$. **c**, Calculated spin state population error (Δz) after Hahn-Echo sequence over relative microwave power error Δg (green) and frequency error Δf (red). **d**, Cumulative noise over sequence length $T_{seq}$ from high to low frequencies of laser (blue), microwave power (green) and microwave frequency (red). Measured deviation of single readout steps $\sigma_1$ indicated by black horizontal line. Varying microwave power is expected to dominate the distribution of magnetic field evaluations on longer timescales. Therefore, results of field evaluations do not share a common central limit over time.

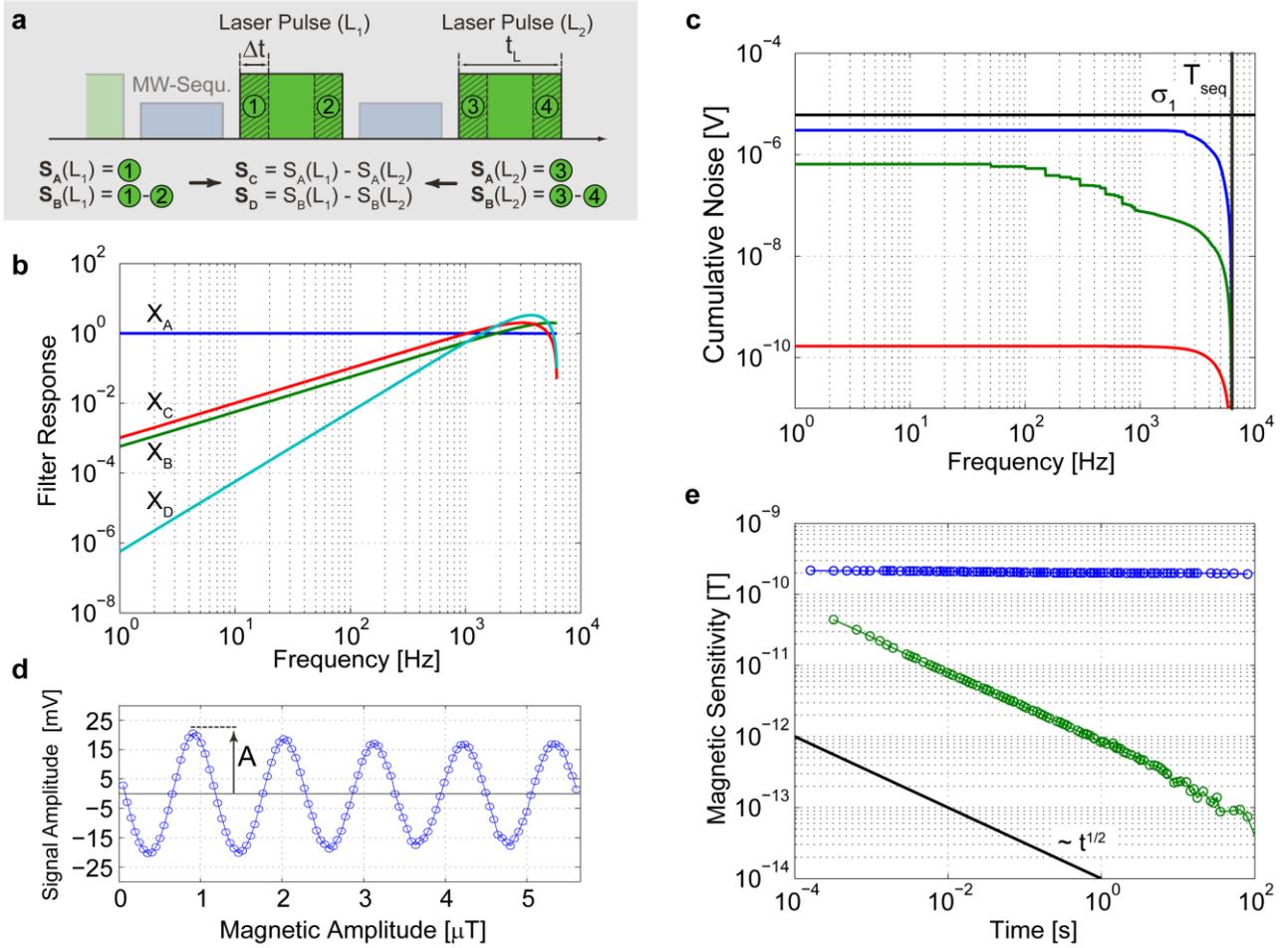

**Fig. 3: Sequence filter and sensitivity scaling. a**, Measurement signals ($S_A - S_D$) represent different ways of taking the signal from the sensor. **b**, Corresponding sequence filter functions resulting from signals $S_A - S_D$. **c**, Cumulative, sequence filtered noise of laser (blue), microwave power (green) and microwave frequency (red) for signal $S_D$. Black, horizontal line indicates single sequence readout deviation ($\sigma_1$). $\sigma_1$ remains the dominant contribution within the timescale shown. **d**, Magnetic measurement of test field with varying field amplitude retrieves sensor response A (equation (1)). **e**, Scaling of magnetic sensitivity (standard deviation over time) of signal $S_B$ (blue) and $S_D$ (green). Slope of black line again indicates the aspired scaling behaviour with $\sqrt{t}$.

14